\documentclass[aps,prb,twocolumn,showpacs,superscriptaddress]{revtex4-1}
\usepackage{amsmath}
\usepackage{amssymb}
\usepackage{bm}
\usepackage[english]{babel}
\usepackage{graphicx}
\usepackage[english]{babel}
\usepackage{color}

\begin{document}

\title{Time-dependent transport of a localized surface plasmon through a linear array of metal
nanoparticles: Precursor and normal mode contributions
}

\author{P.~J.~Compaijen}
\email[]{jaspercompaijen@gmail.com}
\affiliation{Zernike Institute for Advanced Materials, University of Groningen, Nijenborgh 4, 9747 AG Groningen, The Netherlands}

\author{V.~A.~Malyshev}
\email[]{v.malyshev@rug.nl}
\affiliation{Zernike Institute for Advanced Materials, University of Groningen, Nijenborgh 4, 9747 AG Groningen, The Netherlands}

\author{J.~Knoester}
\email[]{j.knoester@rug.nl}
\affiliation{Zernike Institute for Advanced Materials, University of Groningen, Nijenborgh 4, 9747 AG Groningen, The Netherlands}

\date{\today}

\begin{abstract}
We theoretically investigate the time-dependent transport of a localized surface plasmon excitation through a linear array of identical and equidistantly spaced metal nanoparticles. Two different signals propagating through the array are found: one traveling with the group velocity of the surface plasmon polaritons of the system and damped exponentially, and the other running with the speed of light and decaying in a power-~law fashion, as $x^{-1}$ and $x^{-2}$ for the transversal and longitudinal polarizations, respectively. The latter resembles the Sommerfeld-Brillouin forerunner and has not been identified in previous studies. The contribution of this signal dominates the plasmon transport at large distances. In addition, even though this signal is spread in the propagation direction and has the lateral dimension larger than the wavelength, the field profile close to the chain axis does not change with distance, indicating that this part of the signal is confined to the array.
\end{abstract}

\pacs{73.21.−b, 73.20.Mf, 78.67.Sc}

\maketitle

\section{Introduction}
\label{sc:pp_introduction}
Recent advances in fabrication techniques have led to an impulse in the application of nanotechnology, in particular, in the development of nano-electronic chips. Importantly, miniaturizing electronics and improving its performance, requires densely packed interconnects. Due to the crosstalk of elements, it is expected that the application of electronic interconnects becomes problematic for dense packing. A significant improvement in performance can be achieved, when the communication is mediated by light, rather than by electrons.
However, since the size of traditional optical devices is limited by the diffraction, there is a large size-mismatch between the electronic and optical components. Potentially, this problem can be solved using plasmonic materials, where the energy is stored in the
collective excitations of free electrons of a metallic structure - plasmons.~\cite{Barnes2003,Ozbay2006,NovotnyHecht2006,Maier2007,Koenderink2015}
Such optical excitations can be confined well below the diffraction limit of light. Depending on a given geometry, these properties can be exploited for the design of nano-scale optical waveguides,~\cite{Bozhevolnyi2006,Lal2007,Fang2015} optical antennas,~\cite{Bharadwaj2009,Novotny2011,Agio2012,AgioAlu2013} and plasmonic sensors.~\cite{Anker2008,Stewart2008}

A well-known example of a plasmonic waveguide is a chain of metal nanoparticles (MNPs), also known as a plasmonic array. Excitation of one of the MNPs will induce a so-called Localized Surface Plasmon (LSP), dipole of which, due to the Coulombic forces, will couple to the neighboring particles in the chain. In this way, an optical signal can propagate through the chain, maintaining the strong energy confinement properties of the LSP. This system has been first introduced by Quinten \textit{et al.}~\cite{Quinten1998} and since then it has been extensively studied and discussed in the literature. The waveguiding properties of plasmons of such a chain have been investigated in detail, considering both the frequency dependence and the propagation distance.~\cite{Brongersma2000,Maier2003,Citrin2004,Citrin2005,Markel2007,Evlyukhin2007,Govyadinov2008,Alu2010,Willingham2011, Ruting2011,Compaijen2013,Rasskazov2014} Furthermore, due to the large optical cross section associated with the LSP, it has become clear that this system can also be used as a nano-antenna, either to localize the excitation~\cite{Hernandez2005,DeWaele2007,Malyshev2008} or to radiate with a sharp directionality,~\cite{Koenderink2009,Liu2010,Bharadwaj2009,Novotny2011,Coenen2011,Munarriz2013} as well as to directionally excite surface plasmon polaritons.~\cite{Lin2013,Pors2014,Yao2015,Compaijen2016}

The environment of the chain plays an important role in the interactions between the particles in the chain and, therefore, the optical
and guiding properties of the chain are significantly altered when it is embedded in a layered medium.~\cite{Evlyukhin2006,Fung2007,Compaijen2013,Dong2013,Compaijen2015} All these properties are contained in the dispersion relations,~\cite{Weber2004,Koenderink2006,Alu2006,Markel2007,Govyadinov2008,Campione2011,Shore2012,Compaijen2015} from which information about the velocity and the lifetime of the plasmonic modes can be derived.

In spite of the fact that plasmonic arrays have been subjected to a large amount of studies, most of these focused on the steady-state response of the system. For applications in communication, it is of crucial importance to also understand time dependent behavior, that is the main goal of the present study.

Metals are known to be strongly dispersive and dissipative materials at optical frequencies. Because of that, the propagation velocity and damping of the signal are highly dependent on the carrier frequency of the pulse and its duration or, in other words, on its Fourier spectrum.~\cite{Maier2003,Govyadinov2008,Rasskazov2014a} When the dispersion is smooth and dissipation is small, an optical signal will propagate at the group velocity $v_{g}= d\omega(q)/dq$, where $\omega(q)$ is the chain's plasmon dispersion relation. However, both conditions are not necessarily met in plasmonic arrays, where the dispersion relations have steep regions and strong dissipation is an inevitable condition to have subwavelength confinement.~\cite{Khurgin2015}

In this paper, we simulate the propagation of a LSP, created on the leftmost MNP by a Gaussian optical pulse, through a long chain of equidistantly spaced spherical MNPs. Surprisingly, we find two propagating signals: one, traveling with the group velocity and damped exponentially, according to the system's dispersion relation, and the other, running with the speed of light in the surrounding medium and decaying in a power-law fashion, as $x^{-1}$ and $x^{-2}$ for the transversal and longitudinal polarizations, respectively. The latter resembles the Sommerfeld-Brilluoin forerunner~\cite{Sommerfeld1914,Brillouin1914,Brillouin1960} and has not been pointed out in the previous studies. The contribution of this signal dominates the plasmon transport at large distances. Globally, the lateral dimension of the forerunner spreads in the propagation direction, however, the field profile close to the chain axis does not change with distance, indicating that this part of the signal remains confined to the array.

The remainder of this paper is organized as follows. In the next section, we introduce the mathematical model that is used for the analytical and numerical calculations. In Sec.~\ref{sc:pp_dispersionrelations}, the dispersion relations are derived and discussed. Section~\ref{sc:pp_results} first treats the frequency domain solutions, followed by the time-dependent propagation simulations. The obtained results are analyzed using an analytical solution based on the Green's function of the system. Additionally, the lateral spread of the precursor signal is calculated in order to get insight into its confinement to the array. Finally, in Sec.~\ref{sc:pp_summary}, we summarize.

\section{Formalism}
\label{sc:pp_formalism}
\begin{figure}[ht!]
\includegraphics{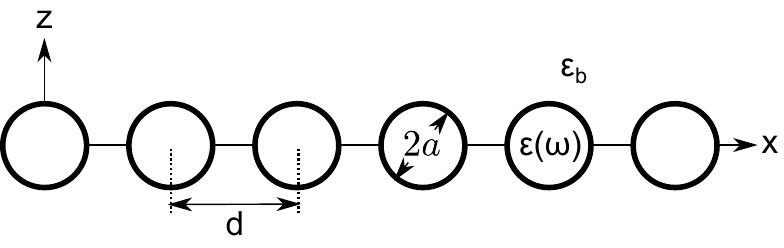}
\caption{\label{fig:pp_geometry} Schematics of the system under consideration: a linear array of identical and equidistantly spaced spherical MNPs with radius $a=25$ nm and center-to-center distance $d=75$ nm. The array is embedded in a medium with permittivity $\epsilon_{b}=2.25$. The chain length $N=4000$. Only the leftmost ($n = 1$) particle is considered to be excited by a light source.}
\end{figure}

The plasmonic waveguide that we consider in this paper is a linear array of identical and equidistantly spaced spherical MNPs. The array is oriented along the x-axis and the center of the leftmost MNP coincides with the origin ($n = 1, x_1 = 0$). The system is depicted schematically in Fig.~\ref{fig:pp_geometry}, in which the center-to-center distance between the MNPs is given by $d=75$ nm and their radii by $a=25$ nm. The array is embedded in a glass matrix with a permittivity $\epsilon_{b}=2.25$. Throughout this work we will consider a chain consisting of $N=4000$ MNPs.
We assume that only the leftmost ($n = 1$) MNP of the array is excited. In practice, this can be achieved with, e.g., a tapered optical fiber, as is used, for example, in near-field optical microscopy.~\cite{NovotnyHecht2006}
Note that the chain length $N = 4000$ is much longer than necessary for any practical application. We make this choice, because it enables us to find the amplitude and velocity of the signal without the complication of reflection from the end of the chain, thereby allowing us to clearly distinguish between different contributions to the signal propagation. In what follows, such distinction is essential.

The frequency dependent response of MNPs is well understood and therefore, we will first calculate the system's optical response in the frequency domain. The MNP's size considered is much smaller than the excitation wavelength and the interparticle spacing satisfies $d\leq3a$. Under these limitations, it suffices to describe the MNPs as oscillating point dipoles.~\cite{Park2004} An external field of the form ${\bf E}\exp(-i\omega t)$ induces a dipole moment in an MNP with an amplitude given by ${\bf p} = \epsilon_b\alpha(\omega){\bf E}$. Here, $\alpha(\omega)$ is the frequency dependent polarizability of the metal, which for spherical MNPs can be written as
\begin{equation}
\label{eq:pp_polarizability}
\frac{1}{\alpha(\omega)}=\frac{\epsilon(\omega)+2\epsilon_{b}}{\epsilon(\omega)-\epsilon_{b}}\frac{1}{a^{3}} - \frac{k_{b}^{2}}{a} - \frac{2i}{3}k_{b}^{3}.
\end{equation}
In this equation, $\epsilon(\omega)$ is the permittivity of the material of which the MNPs are composed and $k_{b}=\sqrt{\epsilon_{b}}\omega/c=\omega/v$ is the wavevector in the background medium, where $c$ and $v$ are the speed of light in vacuum and in the background medium, respectively. The first term in Eq.~(\ref{eq:pp_polarizability}) is the electrostatic polarizability, while the second and third terms account for spatial dispersion and radiation damping, respectively.~\cite{footnote} In this paper, we consider silver MNPs with a permittivity given by the generalized Drude formula:~\cite{Koenderink2007} $\epsilon(\omega)= \epsilon_\infty - \eta\omega_p^2/(\omega^{2} + i\gamma\omega)$ with $\epsilon_\infty = 5.45$, $\eta = 0.73$,  $\omega_p = 16.2\cdot10^{16}$s$^{-1}$, and $\gamma = 8.35\cdot10^{13}$s$^{-1}$, which fits well the tabulated optical data.~\cite{Palik1985}

The induced dipole moment of the MNP will oscillate along with the applied external field and therefore, it will also generate an electric field. The electric field at position $\mathbf{r}$ produced by an oscillating point dipole $\mathbf{p'}$ located at $\mathbf{r'}$ can be written in terms of the Green's tensor $\mathbf{\hat{G}}$ as
\begin{equation}
\label{eq:pp_dipolefield}
\mathbf{E}(\mathbf{r})=\frac{k_{b}^{2}}{\epsilon_{b}}\mathbf{\hat{G}}(\mathbf{r},\mathbf{r'})\mathbf{p'}\ ,
\end{equation}
where the $\omega$-dependence is dropped. Taking into account that the dipole moment of an MNP is proportional to the total field acting on it, which is the sum of the external field and the fields produced by all the other MNPs, we can write the following equation of motion for the dipole moment amplitudes of the coupled system in the frequency domain:
\begin{equation}
\label{eq:pp_eom}
\frac{1}{\epsilon_{b}} \sum_{m(\neq n)} \Big[ \frac{1}{\alpha} \delta_{nm} \mathbf{\hat{I}} - k_{b}^{2} \mathbf{\hat{G}}(\mathbf{r_{n}},\mathbf{r_{m}})\Big] \mathbf{p_{m}} = \mathbf{E_{n}^{\textnormal{ext}}}.
\end{equation}

The exact form of the Green's tensor is strongly dependent on the environment of the dipoles. In the present situation, we consider a chain of MNPs in an otherwise homogeneous background with permittivity $\epsilon_b$. In this case, the Green's tensor is simply given by~\cite{NovotnyHecht2006}
\begin{equation}
\label{eq:pp_G0tensor}
\mathbf{\hat{G}}(\mathbf{r_{n}},\mathbf{r_{m}}) = \Big[\mathbf{\hat{I}}+\frac{\nabla \nabla}{k_{b}^{2}}\Big] \frac{e^{ik_{b}|\mathbf{r}_{n}-\mathbf{r}_{m}|}}{|\mathbf{r}_{n}-\mathbf{r}_{m}|}.
\end{equation}

From the symmetry of the system (see Fig.~\ref{fig:pp_geometry}), it is obvious that only two independent polarizations will exist:
longitudinal ($||$), when the dipole moments are oriented along the chain axis, and transversal ($\perp$), with the dipole moments oriented perpendicular to the chain axis. Therefore, only the diagonal elements of the Green's tensor will be nonzero and we can simplify Eq.~(\ref{eq:pp_G0tensor}) to
\begin{widetext}
\begin{subequations}
\label{eq:pp_greens}
\begin{equation}
\label{eq:pp green longitudinal}
G_{||}(x_{n},x_{m}) = 2\left(-\frac{i}{k_{b}|x_{n}-x_{m}|^{2}} + \frac{1}{k_{b}^{2}|x_{n}-x_{m}|^{3}}\right)e^{ik_{b}|x_{n}-x_{m}|}\ ,
\end{equation}
\begin{equation}
\label{eq:pp green transversal}
G_{\perp}(x_{n},x_{m}) = \left(\frac{1}{|x_{n}-x_{m}|} + \frac{i}{k_{b}|x_{n}-x_{m}|^{2}} - \frac{1}{k_{b}^{2}|x_{n}-x_{m}|^{3}}\right)e^{ik_{b}|x_{n}-x_{m}|} \ ,
\end{equation}
\end{subequations}
\end{widetext}
where we introduced the notations: $G_{||}=G_{xx}$ and $G_{\perp}=G_{yy}=G_{zz}$.

The optical response of the array to a given external excitation can now easily be found by inserting Eq.~(\ref{eq:pp_greens}) into Eq.~(\ref{eq:pp_eom}) and solving the latter for $\mathbf{p_{n}}$. Note that this is the solution in the frequency domain and the external excitation is considered to be a continuous wave of a fixed frequency $\omega$.

In order to investigate how an LSP, created on the leftmost MNP, propagates through the array, we have to solve the time-domain problem. The excitation pulse considered in this paper is assumed to have a Gaussian envelope with a standard deviation $\Delta t$, a
carrier frequency $\omega_{0}$, and is centered around $t=t_{0}$, i.e.,
\begin{equation}
\label{eq:pp_gaussiant}
\mathbf{\tilde{E}}(t)=\mathbf{E_{0}}e^{-i\omega_{0}t}e^{-\left(\frac{t-t_{0}}{\Delta t}\right)^{2}}.
\end{equation}
To probe the effects of dispersion on the LSP propagation, the temporal width $\Delta t$ of the excitation is considered relatively long, so that the spectrum of the pulse is narrow compared to the frequency scale over which the dispersion relation varies strongly.

The response of the system in the time domain, $\mathbf{\tilde{p}_{n}}(t)$, can be obtained by taking the inverse Fourier transform of the
frequency-domain solution $\mathbf{p_{n}}(\omega)$. To this end, we first have to perform the Fourier transform of the above defined Gaussian, Eq.~(\ref{eq:pp_gaussiant}), which will be again a Gaussian
\begin{equation}
\label{eq:pp_gaussianw}
\mathbf{E}(\omega)=\frac{\mathbf{E_{0}}\Delta t}{\sqrt{2}} e^{i(\omega-\omega_{0})t_{0}} e^{-\left(\frac{\omega-\omega_{0}}{2/\Delta t}\right)^{2}} \ .
\end{equation}
Then, substituting ${\bf E}(\omega)$ into Eq.~(\ref{eq:pp_eom}) and solving for the dipole moment per frequency component, $\mathbf{p_{n}}(\omega)$, the time-dependent response $\mathbf{\tilde{p}_{n}}(t)$ can be easily found by taking the inverse Fourier transform, i.e.,
\begin{equation}
\label{eq:pp_ift}
\mathbf{\tilde{p}_{n}}(t)=\frac{1}{\sqrt{2\pi}} \int_{-\infty}^{\infty} d\omega \, \mathbf{p_{n}}(\omega) e^{-i\omega t}  \ .
\end{equation}

\section{Dispersion relations}
\label{sc:pp_dispersionrelations}
In any propagation problem, the normal mode representation provides a convenient formalism to get insight into details of the system's dynamics. Thus, finding the normal modes of the plasmonic chain is of crucial importance. They are found by solving Eq.~(\ref{eq:pp_eom}) for $\mathbf{E}=\mathbf{0}$. Even though several schemes are available to find the normal modes, in general, this is not a straightforward task. For short arrays, the summation in Eq.~(\ref{eq:pp_eom}) can be carried out directly and the modes are easily found using the method proposed by, e.g., Weber and Ford.\cite{Weber2004} Due to poor convergence of the summation, it becomes increasingly difficult to apply this method when longer arrays of MNPs are considered.

Interestingly, the infinite chain approximation has been shown to be already accurate for relatively short chains.\cite{Citrin2005} In this case, the sums in Eq.~(\ref{eq:pp_eom}) can be rewritten in terms of polylogarithms, which can be evaluated using analytical continuation.~\cite{Weber2004,Citrin2006,Alu2006} For an infinite periodic chain with an inter-particle spacing $d$, we look for the collective modes of the system in the form of Bloch waves, i.e., $\mathbf{p_{m}} = \mathbf{p}\,\textnormal{exp}[iqmd]$. Substituting this in Eq.~(\ref{eq:pp_eom}) with $\mathbf{E}=\mathbf{0}$ yields
\begin{equation}
\label{eq:pp_eombloch}
\left[\frac{1}{\alpha}-\sum_{m (\neq n)}k_{b}^{2} G_{\beta}(nd,md) e^{iq(m-n)d}\right]p=0,
\end{equation}
where the Green's tensor is replaced by $G_{\beta}$, the scalar interaction for either longitudinal or transversal polarization, i.e., $\beta=||$ or $\perp$, respectively. The second term in the square brackets is the so-called dipole sum, $S_{\beta}(k_{b},q)$. Splitting the summation in two parts, corresponding to $m > n$ and $m < n$, i.e., forward and backward interactions, $S_{\beta}(k_{b},q)$ can be rewritten as
\begin{widetext}
\begin{subequations}
\begin{equation}
\label{eq:pp_dipolesum longitudinal}
S_{||}(k_{b},q)=\frac{1}{d^{3}}\left(-2ik_{b}d \left[\textnormal{Li}_{2}(z^{+})+\textnormal{Li}_{2}(z^{-})\right] + 2\left[\textnormal{Li}_{3}(z^{+})+\textnormal{Li}_{3}(z^{-})\right]\right)\ ,
\end{equation}
\begin{equation}
\label{eq:pp_dipolesum transversal}
S_{\perp}(k_{b},q)=\frac{1}{d^{3}}\big(k_{b}^{2}d^{2} \left[\textnormal{Li}_{1}(z^{+})+\textnormal{Li}_{1}(z^{-})\right] + ik_{b}d\left[\textnormal{Li}_{2}(z^{+})+\textnormal{Li}_{2}(z^{-})\right] \\
+ \left[\textnormal{Li}_{3}(z^{+})+\textnormal{Li}_{3}(z^{-})\right] \big) \ ,
\end{equation}
\end{subequations}
\end{widetext}
where $\textnormal{Li}_{s}(z)=\sum_{n=1}^{\infty}z^{n}/n^{s}$ is the polylogarithm of order $s$ (see Ref.~\onlinecite{Lewin1981}) and the arguments $z^{\pm}=\textnormal{exp}[i(k_{b}\pm q)d\,]$.

In this approximation, the dispersion relation $\omega(q)$ can be obtained by numerically solving the following equation
\begin{equation}
\label{eq:pp_dispersion}
\frac{1}{\alpha(\omega)}-S_{\beta}(k_{b},q)=0.
\end{equation}
\begin{figure}
\includegraphics{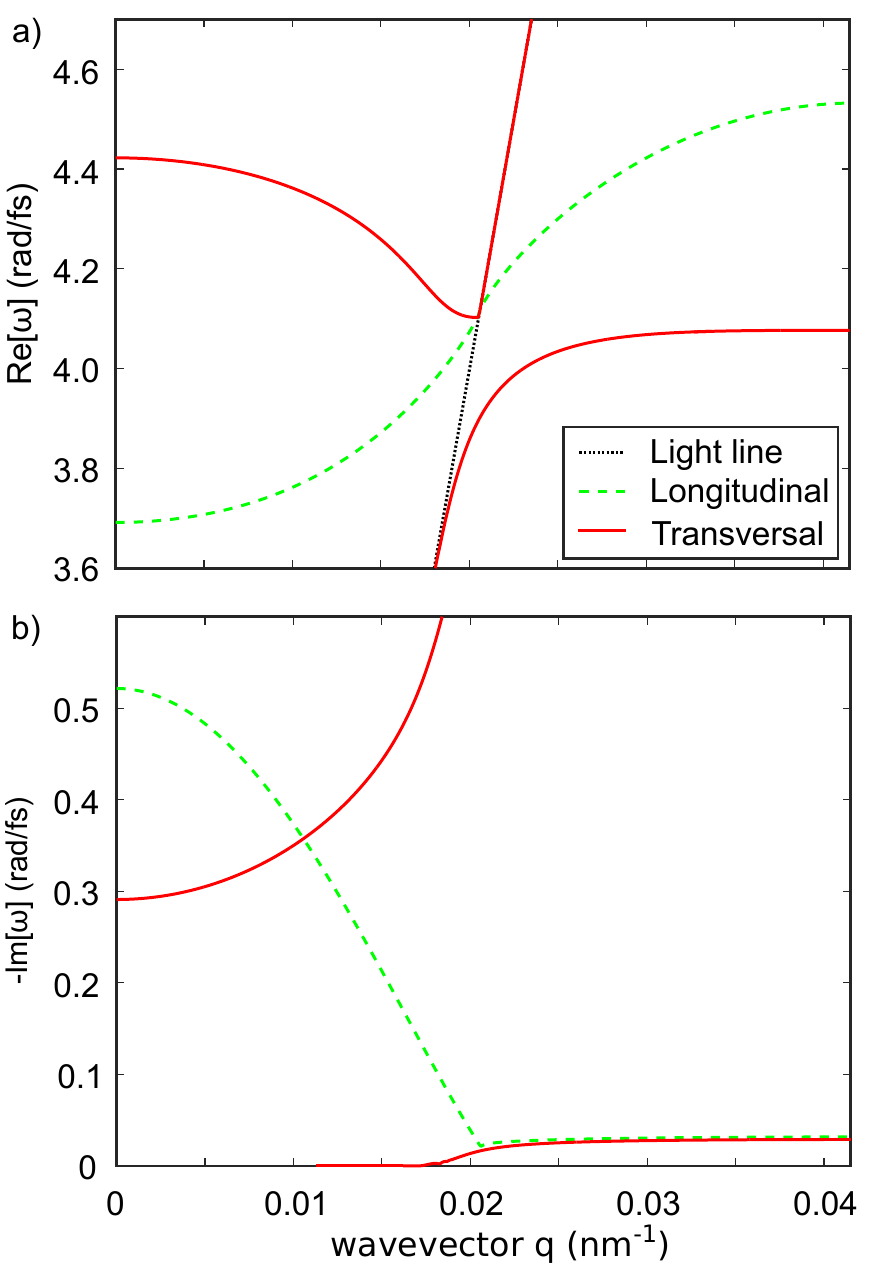}
\caption{\label{fig:pp_dispersionrelations} The dispersion relations $\omega(q)$ of an infinite chain of silver spherical MNPs of radius $25$ nm and spacing of $75$ nm in a glass background medium. a) Real part of the frequency $\omega$ versus the wavevector $q$ for longitudinally (green) and transversally (red) polarized plasmons. The black dotted line indicates the dispersion of light in glass. b) The imaginary part of the frequency $\omega$ versus the wavevector $q$. For reference purposes: the plasmon resonance of an individual MNP is $\omega_{LSP}=4.14-0.19i$ rad/fs. Note that the wavevector $q$ runs from $0$ to the first Brillouin zone edge at $\pi/d$.}
\end{figure}

\noindent
Figure~\ref{fig:pp_dispersionrelations} shows the thus obtained dispersion relations for an infinite linear array of silver nanospheres with the parameters as in Fig.~\ref{fig:pp_geometry}; these results are in agreement with previous calculations of the dispersion relations of linear plasmonic arrays.~\cite{Weber2004,Citrin2005,Koenderink2006} The upper and lower panel show the real and imaginary parts of the normal mode frequencies, respectively. A time-dependence of $\textnormal{exp}(-i\omega t)$ is assumed, and therefore, the real part of the frequency $\textnormal{Re}[\omega(q)]$ refers to the oscillation frequency, whereas the imaginary part $\gamma(q) = \textnormal{Im}[\omega(q)]$ gives rise to exponential damping in time. In fact, $\gamma(q)^{-1}$ determines the mode's lifetime.

From Fig.~\ref{fig:pp_dispersionrelations}a one can see a clear difference between the longitudinal and transversal chain modes. Firstly, at $q = 0$ the longitudinal mode reaches a minimum that lies below the plasmon resonance of a single MNP, whereas the transversal mode reaches a maximum, lying above the single particle resonance. This reflects the fact that the sign of the near-field dipole-dipole coupling is opposite for these two modes [see Eqs.~(\ref{eq:pp green longitudinal}) and~(\ref{eq:pp green transversal})].
Secondly, an important difference between both polarizations is the anti-crossing at the light line that is only observed for the transversal case, indicating polariton behavior of the coupled plasmon-photon system.
This results from the fact that light is a transversal wave and therefore couples strongly to transversally polarized plasmons that propagate along the array, giving rise to the avoided crossing at $\textnormal{Re}[\omega(q)] = vq$. Such a coupling does not exist for longitudinal plasmons. This is also reflected in the fact that the radiative interaction term, proportional to $1/x_n$, is absent for the longitudinal interaction, whereas it is present for the transversal polarization.

Looking at Fig.~\ref{fig:pp_dispersionrelations}b, one observes a large reduction of the losses when the light line is crossed,
$k_{b} = \textnormal{Re}[\omega]/v_p$. Within the light cone, $q < k_{b}$, the modes of the chain suffer from both ohmic and radiative losses. However, modes with $q > k_{b}$ cannot couple to the free-space radiation and only will suffer from Ohmic losses. These are the so-called guided modes of the system and the ones of interest for signal propagation.
The modes inside the light cone ($q < k_b$) are characterized by significant losses, predominantly of radiative nature. These are the so-called leaky modes of the chain. Their contribution to the LSP transport is marginal as compared with the guided modes.

From the dispersion relations, one can derive important properties of the energy transport. Each mode propagates at its own phase velocity, defined as $v_{p}(q) = \textnormal{Re}[\omega(q)]/q$. When a collection of modes (a wavepacket) is excited with a carrier frequency $\omega_0$, this will propagate with the group velocity, given by $v_g(q) = \frac{d}{dq}\textnormal{Re}[\omega(q)]|_{\omega_{0}}$.
This allows one to determine the propagation length $l(q)$ for the normal mode, deriving it from the relation $l(q) = v_g(q)/\gamma(q)$, as was done in Ref.~\onlinecite{Koenderink2006}.

\section{Signal propagation}
\label{sc:pp_results}
\subsection{Frequency domain}
\label{ssc:pp_resultsfreqdomain}
\begin{figure}
\includegraphics{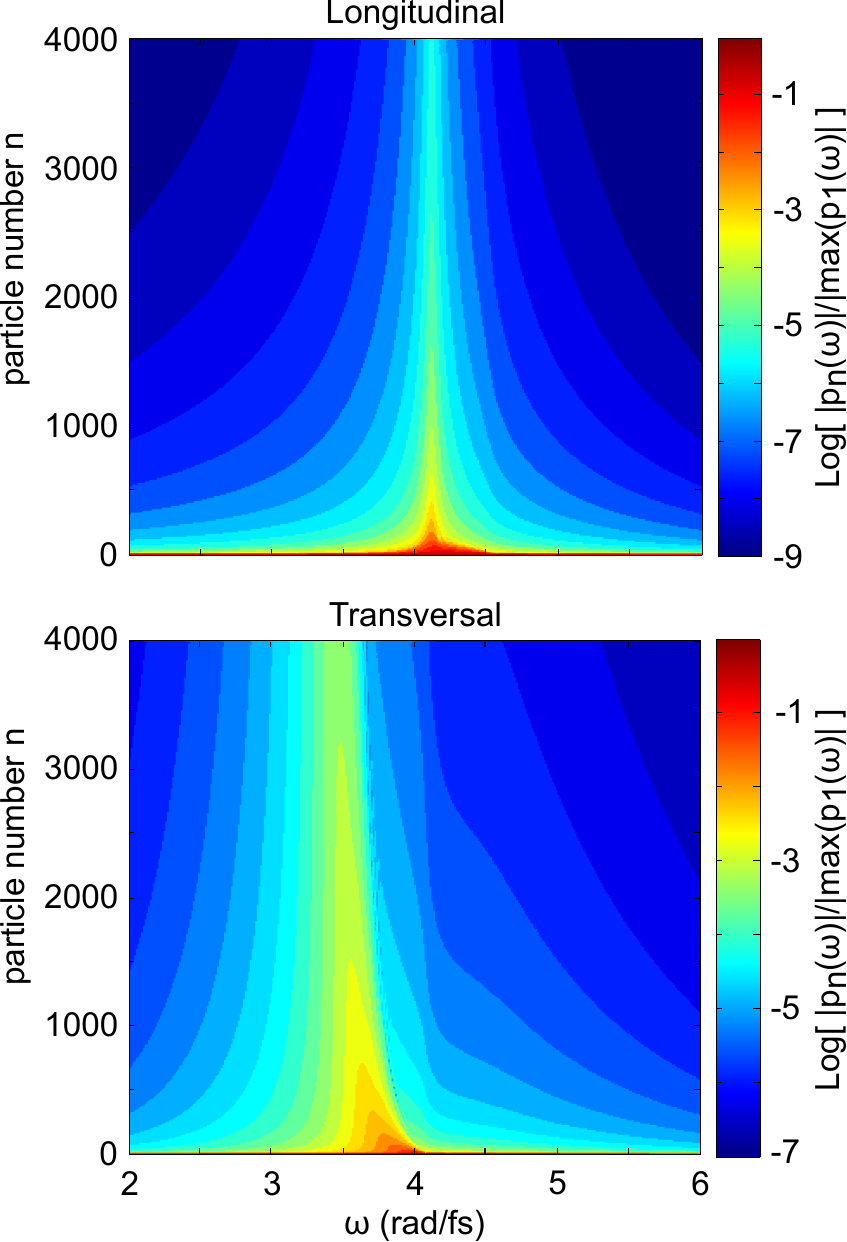}
\caption{\label{fig:pp_freqdomain} The frequency domain characterization of the energy transport through a plasmonic array
comprising $4000$ silver nanospheres with radii $a=25$ nm and interparticle spacing $d=75$ nm.
The data were obtained by numerically solving Eq.~(\ref{eq:pp_eom}) as a function of frequency $\omega$ under the condition of excitation of the leftmost ($n = 1$) particle of the array.
The magnitudes of the dipole moments' Fourier components $\{\mathbf{p_{n}}\}$ are plotted on a logarithmic scale. The dipole moment of each particle is normalized with respect to the maximum magnitude of the dipole moment of the leftmost ($n = 1$) particle.
The maximum absolute value of the dipole moment of the last particle occurs at $\omega=4.12$ rad/fs and at $\omega=3.46$ rad/fs for the longitudinal and transversal polarizations, respectively.
}
\end{figure}
\begin{figure}
\includegraphics{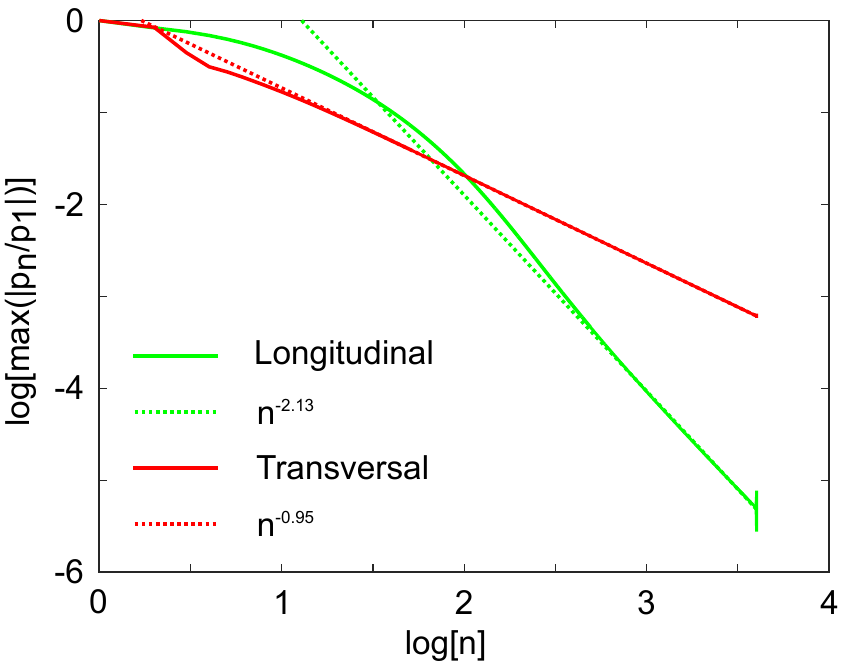}
\caption{\label{fig:pp_freqdomain_decay} Double logarithmic plot of the maximum magnitude of the dipole moment $\mathbf{p_n}$ for each particle, obtained from Fig.~\ref{fig:pp_freqdomain}, as a function of the position $n = x_n/d$ along the chain axis.
The dipole moments are normalized with respect to the maximum magnitude of the leftmost ($n = 1$) dipole. It is seen that after an initial exponential stage, the curves become linear, indicating a power-law decrease of $\mathbf{p_n}$ along the chain axis. Using a linear fitting, it is
found that the decay follows the power laws $n^{-2.13}$ and $n^{-0.95}$ for the longitudinal and transversal polarizations, respectively.
}
\end{figure}
In Sec.~\ref{sc:pp_formalism}, the polarizability $\alpha(\omega)$ of a single MNP and the interaction between two MNPs $\mathbf{\hat{G}}(\mathbf{r_{n}},\mathbf{r_{m}})$ where given in the frequency domain. The optical response of the chain can now easily be obtained by inserting these quantities in Eq.~(\ref{eq:pp_eom}) and solving it numerically. The frequency domain response of the
system is depicted in Fig.~\ref{fig:pp_freqdomain} for a chain consisting of $N=4000$ silver nanospheres with radii $a=25$ nm and
interparticle spacing $d=75$ nm. As before, the excitation source is considered to act only on the leftmost ($n = 1$) particle. For both the longitudinal and transversal polarizations, it is clearly seen that the dipole moments' magnitudes are large close to the starting point, but gradually decrease along the chain. Figure~\ref{fig:pp_freqdomain} shows that more towards the end of the array, an efficient energy transmission occurs only in a narrow frequency interval. Furthermore, while at the beginning of the chain the dipole moments' magnitudes are larger for the longitudinal polarization, towards the chain end, the transport is more efficient for the transversal polarization. To elucidate this effect better, a graph showing the maximum values of the contour plots of Fig.~\ref{fig:pp_freqdomain} as a function of the distance along the chain is given in Fig.~\ref{fig:pp_freqdomain_decay}. From this
figure, one observes that indeed for the initial part of the chain, the longitudinal polarized transport dominates over the transversal one, but after about 130 particles the situation is opposite.
Interestingly, whereas all collective modes damp exponentially due to dissipation, the line of maximum energy transport at large distances follows a power law , i.e., $|\mathbf{p_n}| \propto x_n^{-s}$. From the slope of the decays in Fig.~\ref{fig:pp_freqdomain_decay}, one can determine the value of $s$, which turns out to be $s\approx 0.95$ and $s\approx 2.13$ for the transversal and longitudinal polarizations, respectively.
%
%

\subsubsection{Discussion of the frequency-domain results}
\label{ssc:pp_resultsfrequencydiscussion}
Even though the data presented in Fig.~\ref{fig:pp_freqdomain} have been calculated by numerically solving Eq.~(\ref{eq:pp_eom}) for a finite chain, in order to shed light on the surprising geometrical decay at long distances, it is informative to consider the analytical solution to Eq.~(\ref{eq:pp_eom}) for an infinite chain. In this case, if only a single particle is excited, the solution to Eq.~(\ref{eq:pp_eom}) represents essentially the Green's function for the dipoles, which is given by the Fourier transform of $\alpha^{-1}(\omega)-S_{\beta}(k_{b},q)$:
\begin{equation}
\label{eq:pp_greensfunction}
P_{\beta}(x_{n},0)=\frac{d}{2\pi\epsilon_{b}} \int_{-\frac{\pi}{d}}^{\frac{\pi}{d}} dq \, \frac{e^{iqx_{n}}}{1/\alpha(\omega)-S_{\beta}(k_{b},q)}\ ,
\end{equation}
where $P(x_{n},0)$ is the dipole moment that is induced in the particle situated at $x_{n}$, when only the particle at $x=0$ is excited.
Note that, in spite the fact that  Eq.~(\ref{eq:pp_greensfunction}) represents the Green's function for an infinite chain, asymptotically (for $x_n \gg d$) it is also a good approximation for a semi-infinite chain: far from the boundary, the influence of the the latter is negligible.

\begin{figure}
\includegraphics{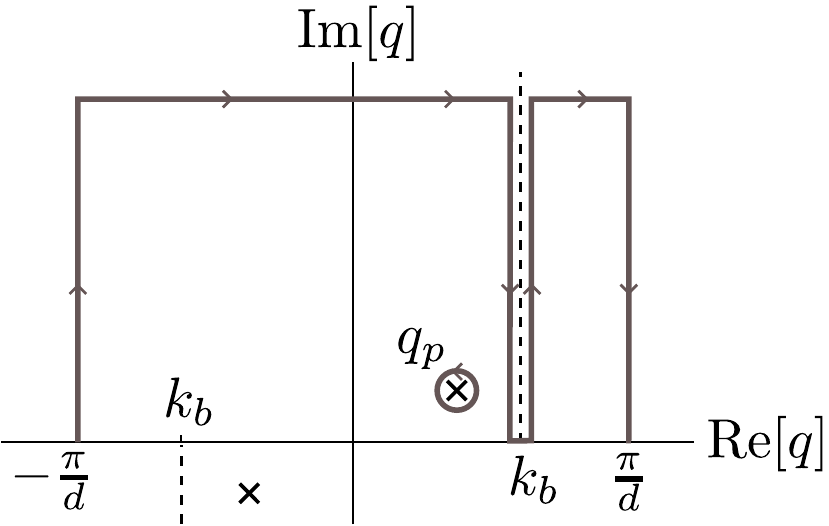}
\caption{\label{fig:pp_integrationpath}
The integration path for evaluating the integral in Eq.~(\ref{eq:pp_greensfunction}). The crosses indicate the position of poles, while the dashed lines denote the logarithmic branch cats.
}
\end{figure}

The integration in Eq.~(\ref{eq:pp_greensfunction}) runs over the real axis from $q=-\pi/d$ to $q=+\pi/d$, which are the edges of the first Brillouin zone. To find the different contributions to the integral, we deform the integration path into the complex plane. In the present case, we can move the path into the upper half-plane, to ensure convergence of the integral for positive values of $x_{n}$. Taking a close look at the integrand in Eq.~(\ref{eq:pp_greensfunction}), we observe that for performing the integration, we have to take into account the poles, given by $1/\alpha(\omega)-S_{\beta}(k_{b},q)=0$, and the logarithmic branch cuts at $q=\pm k_{b}$, originating from the polylogarithms in the dipole sum $S_\beta$. The resulting integration contour is shown in Fig.~\ref{fig:pp_integrationpath}.

Since the integrand in Eq.~(\ref{eq:pp_greensfunction}) is identical on both Brillouin zone edges, the contributions coming from the left and right vertical parts of the integration contour cancel. In addition, the integrand is exponentially decaying as a function of the imaginary part of $q$. Therefore, the contribution of the upper part of the contour tend to zero as long as we move the path far enough away from the real axis. Interestingly, this shows that there are two contributions to the dipole moment of the particle at $x_{n}$, one coming from the pole and the other from the branch cut. The poles of the integrand in Eq.~(\ref{eq:pp_greensfunction}) correspond exactly to the collective modes of the system. The above conclusion therefore implies that the response of the system cannot be simply written as a superposition of the modes, the contribution of the branch cut also has to be taken into account.

The contribution of the pole, $P^{pole}_\beta$, can easily be calculated using the Residue Theorem. Denoting the position of the pole by $q=q_{p}$, we find
\begin{equation}
\label{eq:pp_pole}
P^{pole}_{\beta}(x_{n},0)=\frac{d}{i\epsilon_{b}}\frac{e^{iq_{p}x_{n}}}{S'_{\beta}(k_{b},q_{p})}\ ,
\end{equation}
where $S'$ denotes the first derivative of $S$ with respect to $q$. Due to the presence of dissipation, one knows that the poles will occur at complex values of $q$. Therefore, it is clear that the contribution, coming from the collective modes of the system, will decay exponentially as a function of $x_{n}$ with a decay constant given by $\textnormal{Im}[q_{p}]$.

The origin of the power law decay of the signal for large $x_{n}$ thus has to come from the contribution of the integration around the branch cut, $P^{bc}_{\beta}$. Since the integrand vanishes exponentially according to $\textnormal{exp}[-\textnormal{Im}[q]x_{n}]$, the contributions of both vertical paths along the cut will be zero for large $x_{n}$. Therefore, at large distances, we expect to only see the contribution from the integration over the real axis from $k_{b}-\epsilon$ to $k_{b}+\epsilon$, for arbitrarily small $\epsilon$, i.e.,
%
\begin{equation}
\label{eq:pp_branchcut1}
P^{bc}_{\beta}(x_{n},0)=\lim_{\epsilon\rightarrow 0} \frac{d}{2\pi\epsilon_{b}} \int_{k_b-\epsilon}^{k_b+\epsilon} dq \,  \frac{e^{iqx_{n}}}{1/\alpha(\omega)-S_{\beta}(k_{b},q)} \ .
\end{equation}
%
To find the asymptotic behavior of this integral for large $x_{n}$, we integrate twice by parts and obtain
%
%
\begin{widetext}
\begin{equation}
\label{eq:pp_branchcut2}
\begin{split}
P^{bc}_{\beta}(x_{n},0) = \lim_{\epsilon \to 0} \frac{d}{2\pi\epsilon_{b}} \left(  \frac{1}{ix_{n}} \frac{e^{iqx_{n}}}{1/\alpha-S_{\beta}} \Big\rvert_{k_b - \epsilon}^{k_b + \epsilon} + \frac{1}{x_{n}^{2}} \frac{e^{iqx_{x}} S'_{\beta} }{\left(1/\alpha-S_{\beta} \right)^{2}} \Big\rvert_{k_b-\epsilon}^{k_b + \epsilon}
- \frac{1}{x_{n}^{2}} \int_{k_b-\epsilon}^{k_b+\epsilon} dq \, \frac{e^{iqx_{n}} [(1/\alpha -S_{\beta})S''_{\beta}-2(S'_{\beta})^{2}]}{\left( 1/\alpha(\omega)-S_{\beta} \right)^{3}} \right).
\end{split}
\end{equation}
\end{widetext}
Note that the dependence of $S_\beta$ on $k_{b}$ and $q$ is suppressed in Eq.~(\ref{eq:pp_branchcut2}).

The branch cut originates from the polylogarithms in the dipole sum $S_{\beta}$. The arguments of the polylogarithms are $z^{+}$ and $z^{-}$, which reduce to $z^{+}=\textnormal{exp}[2ik_{b}]$ and $z^{-}=\textnormal{exp}[\mp i\epsilon]$ at the integration boundaries. This implies that the branch cut at $q=+k_{b}$ is only due to the polylogarithms containing $z^{-}$, i.e the forward interactions. One can show that $\lim_{\epsilon\rightarrow 0} \left[\textnormal{Li}_{s}(e^{+i\epsilon}) - \textnormal{Li}_{s}(e^{-i\epsilon})\right]$ equals $0$ for $s=2$ and $3$, but will be nonzero for $s=1$. For transversal polarization, polylogarithms with $s=1,2$ and $3$ occur. Therefore, none of the terms in Eq.~(\ref{eq:pp_branchcut2}) vanish, indicating that the large-distance response of the system decays as $1/x_n$, in close agreement to what was observed in Fig.~\ref{fig:pp_freqdomain_decay}. For longitudinal polarization, only second and third order polylogarithms occur in the dipole sum, and therefore $S_{||}(k_{b},k_{b}+\epsilon)=S_{||}(k_{b},k_{b}-\epsilon)$ as $\epsilon$ goes to zero, implying that the first term in Eq.~(\ref{eq:pp_branchcut2}) disappears. The second term of Eq.~(\ref{eq:pp_branchcut2}) contains the first derivative of $S$ with respect to $q$. Using the fact that the derivative of a polylogarithm with respect to the argument decreases its order by one, we deduce that $S'_{||}$ contains a polylogarithm of the first order and therefore, the second term does not vanish for arbitrarily small $\epsilon$. Thus, the large-distance response for longitudinal polarization decays as $1/x_n^{2}$, in close agreement to what was also observed in Fig.~\ref{fig:pp_freqdomain_decay}.

\subsection{Time domain}
\label{ssc:pp_resultstimedomain}
As was mentioned before, the time-dependent propagation can easily be obtained from the frequency domain solution with the aid of the inverse Fourier transform. The Gaussian pulse with a carrier frequency of $\omega_{0}$ has a Gaussian spectral distribution centered around $\omega_{0}$, as is given by Eqs.~(\ref{eq:pp_gaussiant}) and~(\ref{eq:pp_gaussianw}). To obtain the frequency domain solution depicted in Fig.~\ref{fig:pp_freqdomain}, the amplitude of the excitation was put equal for all frequencies. Since the external excitation enters linearly in the equations, the response to a Gaussian excitation can be obtained by simply multiplying the solution with a Gaussian spectral distribution. Numerically this is advantageous, since the time-consuming frequency domain response only has to be calculated once and consequently, the frequency domain response of the chain to a Gaussian excitation for different values of $\omega_{0}$ can be generated very fast. The inverse Fourier transform was performed using the build-in Fast Fourier Tranform function of Matlab R2015a. To obtain a sufficiently high resolution in the time-domain, the frequency grid that is used had a step size of $0.001$ rad/fs. Furthermore, the considered pulse was chosen relatively long compared to the oscillation period of the normal modes, with a width of $\Delta t=80$ fs. This implies that the spectral distribution has width of $\Delta \omega=0.025$ rad/fs, which is narrow enough to probe the influence of different $\omega_{0}$.

\begin{figure}[ht!]
\begin{center}
\includegraphics{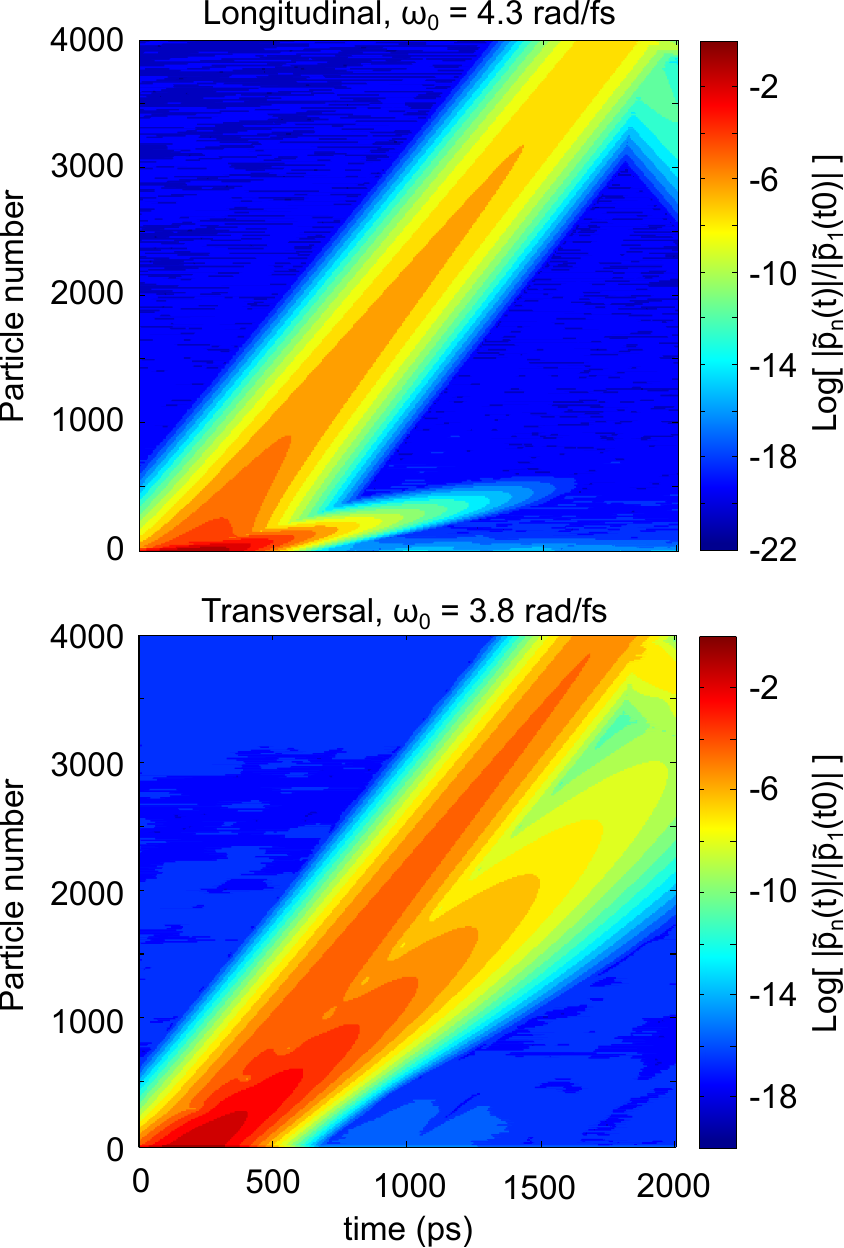}
\caption{\label{fig:pp_prop_contour} The time-dependent response of a plasmonic array consisting of $4000$ silver nanospheres with radiaii $a=25$ nm and a center-to-center spacing $75$ nm. The magnitude of the dipole moment $\mathbf{\tilde{p}_{n}(t)}$ is plotted as a function of the particle number and time.
Only the leftmost ($n = 1$) particle of the chain is excited by a Gaussian pulse of a width (standard deviation) $\Delta t=80$ fs and a carrier frequency $\omega_{0}=4.3$ rad/fs and $\omega_{0}=3.8$ rad/fs for the longitudinal and transversal polarization, respectively.
The dipole moments are normalized to the maximum magnitude of the leftmost dipole, $\mathbf{\tilde{p}_{1}(t_0)}$.}
\end{center}
\end{figure}

An example of the time-dependent pulse propagation through the considered system is given in Fig.~\ref{fig:pp_prop_contour}, where the absolute value of the dipole moment, $|\mathbf{\tilde{p}_{n}(t)}|$, is plotted as a function of the particle number and time. For longitudinal polarization, the carrier frequency $\omega_{0}$ is $4.3$ rad/fs, while for transversal polarization $\omega_{0}=3.8$ rad/fs. Note that these values are just chosen for illustration purposes and that calculations have been performed for a wide range of $\omega_{0}$. The propagation velocity can be determined from the slope of the contours. For both polarizations, one can clearly distinguish two different signals: a fast, slowly decaying signal over the diagonal, and a slower, more lossy signal. In the top right corner the reflection of the fast signal at the end of the chain can be seen. Using Brillouin's definition of signal velocity $v_{s}$,~\cite{Brillouin1960} the slope of the contours was determined by tracing the maximum value of the signal as a function of time. The velocity of the fast signal is found to be exactly equal to the velocity of light in the embedding glass matrix, $200$ nm/fs, and, interestingly, is independent of the carrier frequency of the exciting pulse, $\omega_{0}$. In contrast, the velocity of the slower signal turns out to be strongly dependent on $\omega_{0}$, and as will be shown, it is this signal that matches the guided plasmon modes of the array.

\subsubsection{Discussion of the time domain results}
\label{ssc:pp_resultstimediscussion}
The observation of two signals propagating with different velocities hints towards the two different contributions that were found in the frequency domain simulations in the Sec.~\ref{ssc:pp_resultsfreqdomain}. Since we have obtained analytical expressions for both contributions, by taking the Fourier transform we can find the velocities at which these signals are propagating. The integrals are complicated to evaluate exactly, however, in order to obtain the propagation velocities, a full calculation is not necessary. The Fourier transform of the contribution coming from the branch cut is given by
\begin{equation}
\label{eq:pp_branch_t}
\tilde{P}^{bc}_{\beta}(x_{n},0,t) = \int d\omega\, P^{bc}_{\beta}(x_{n},0,\omega)\, e^{i [k_{b}(\omega) x_{n}- \omega t]} \ ,
\end{equation}
which is a superposition of plane waves $\textnormal{exp}[i(k_{b}(\omega)x_{n}-\omega t)]$ with different frequencies. The functional dependence of $k_{b}$ on $\omega$ is given by $k_{b}=\omega/v$. Therefore, it is evident that this contribution propagates with the velocity $v$, i.e. the speed of light in the surrounding medium, independently of the excitation frequency. Thus, the fast propagating signal arises from the integration along the logarithmic branch cut.

Similarly, for the contribution coming from the pole of Eq.~(\ref{eq:pp_greensfunction}), i.e. from the collective plasmons of the chain, we can write
\begin{equation}
\label{eq:pp_pole_t}
\tilde{P}^{p}_{\beta}(x_{n},0,t) = \int d\omega\, P^{p}_{\beta}(x_{n},0,\omega) e^{i [q_{p}(\omega) x_{n} - \omega t]} \ .
\end{equation}
Although we do not have a simple analytical expression for $q_{p}(\omega)$, to a first approximation, we can write $q_{p}(\omega)=q_{p,0}+\frac{dq_{p}}{d\omega}\big\rvert_{\omega_{0}}(\omega-\omega_0)$. Inserting this in the above equation, we find that the signal is propagating according to $\textnormal{exp}[i(\frac{dq_{p}}{d\omega}\big\rvert_{\omega_{0}} x_{n} - \omega t)]$, i.e. the propagation velocity is $\frac{d\omega}{dq_{p}}\big\rvert_{\omega_{0}}$. Within this approximation, this is equal to the group velocity $v_g$ at the carrier frequency $\omega_{0}$. Therefore, the collective plasmons of the chain propagate with the group velocity, which can be derived from the dispersion relation.

\begin{figure}[ht!]
\begin{center}
\includegraphics{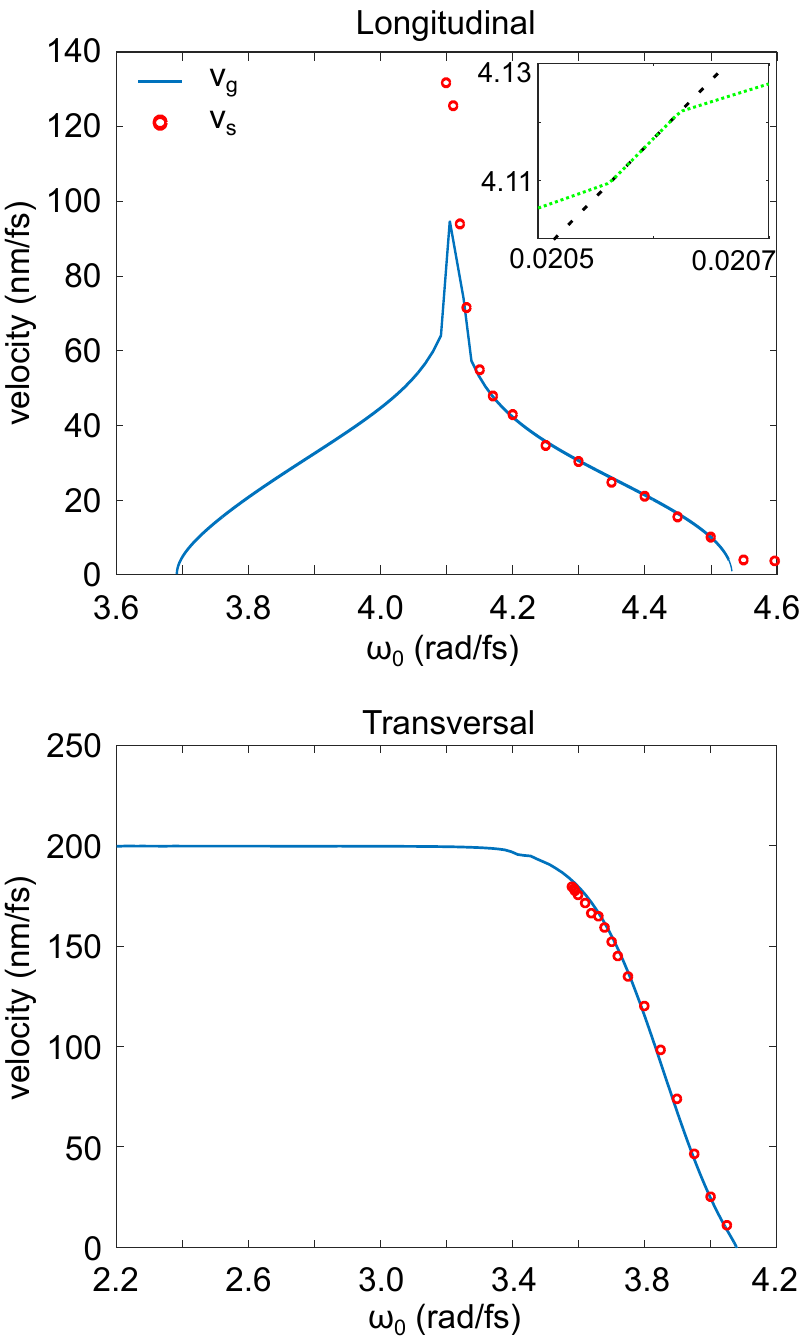}
\caption{\label{fig:pp_velocities} Comparison of the group velocity $v_{g}$ (solid curve), calculated from the dispersion relation $\omega(q)$, and the signal velocity $v_{s}$ (open circles), obtained from the time-dependent propagation shown in Fig.~\ref{fig:pp_prop_contour}. The inset shows a close-up of the crossing point of the dispersion curve for longitudinal polarization with the light line.}
\end{center}
\end{figure}

To verify the above result, Fig.~\ref{fig:pp_velocities} shows both the group velocity, derived from the dispersion relation, and the signal velocity $v_{s}$ of the slower signal, which is obtained by tracing the maximum in the time-dependent propagation simulations. A very close match between both velocities is observed, confirming the hypothesis. Close to the light line the comparison becomes worse. This is because near the light line the changes in the dispersion relation are large, and therefore, taking only the first order in the Taylor series is not sufficient, but higher order terms have to be taken into account.

For longitudinal polarization at high frequencies we still observe a signal, whereas according to the dispersion relation no modes exist for this frequency. However, due to the width of the excitation and the broadening of the modes arising from dissipation, there is still sufficient overlap to excite the modes at the edge of the first Brillouin zone.

Surprisingly, for longitudinal polarization, relatively large group and signal velocities are recorded, much larger than expected from looking at the slope of the dispersion curve in Fig.~\ref{fig:pp_dispersionrelations}. As is shown in the inset of Fig.~\ref{fig:pp_velocities}, in which a close-up of the light line crossing for longitudinal polarization is plotted, the dispersion relation coincides with the light line for a small selection of wavevectors. This implies that for a very narrow frequency interval, the plasmons will actually propagate with the velocity of light. This is the reason for the peak around $\omega_{0}=4.1$ rad/fs for longitudinal polarization. The above result might seem counterintuitive, because radiative interaction only exists for transversal polarization. However, in the interaction there is also the non-static intermediate-field term, proportional to $x_n^{-2}$, that results from retardation and gives rise to a cusp in a narrow region around the crossing point $\omega(q) = vq$.~\cite{Citrin1992,Gartstein2007}

\begin{figure}[ht!]
\begin{center}
\includegraphics{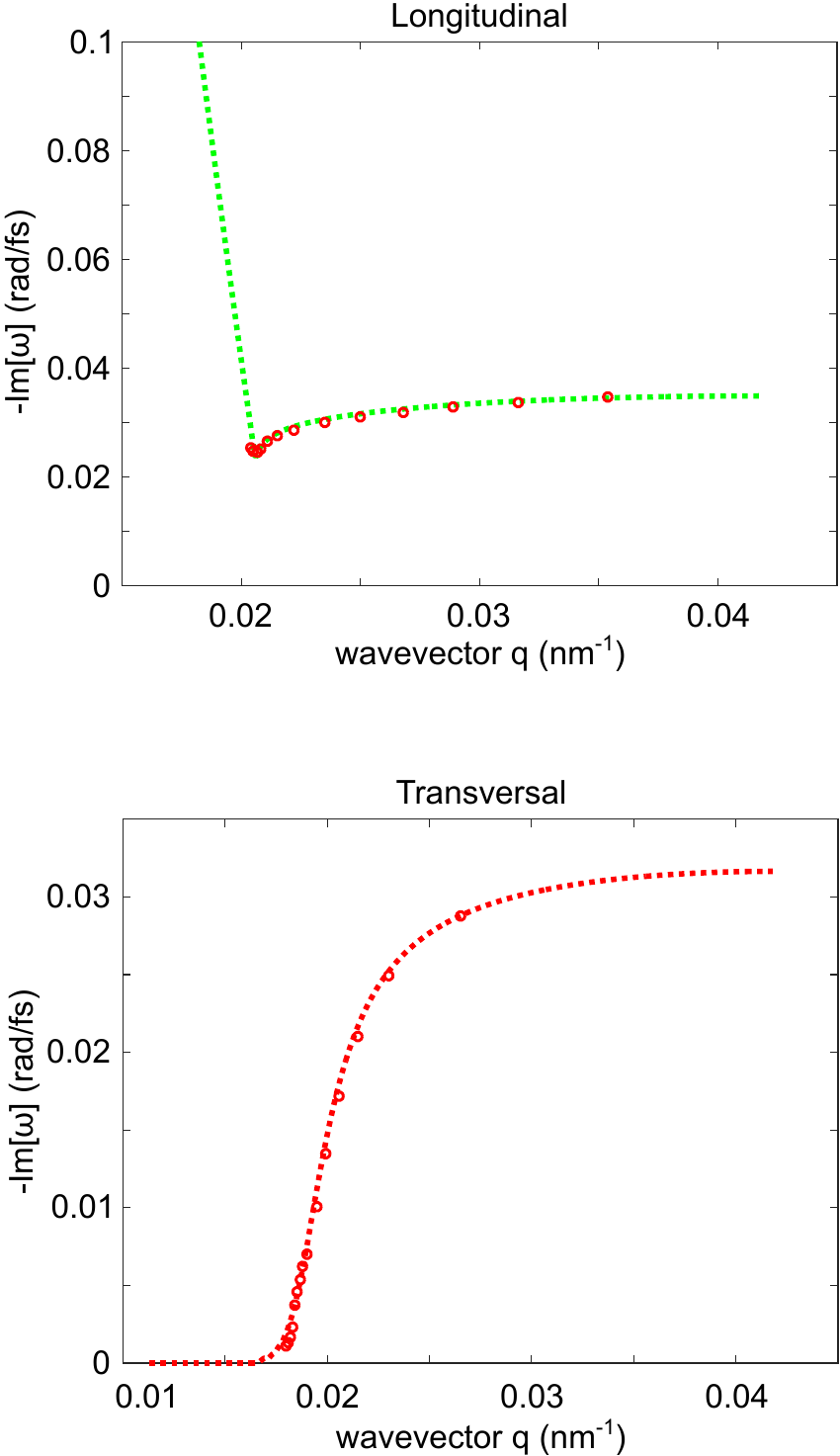}
\caption{\label{fig:pp_lifetimes} The lifetime of the normal modes, longitudinal (upper panel) and transversal (lower panel), respectively. The dashed lines represent the imaginary part of the normal mode frequencies, $1/\textnormal{Im}[\omega(q)]$, as a function of the wavevector $q$. The open circles show the decay constants that are obtained by tracing the maximum amplitude of the slower signal in the pulse propagation simulations for different carrier frequeces $\omega_0$.}
\end{center}
\end{figure}

In addition to the signal velocities, we can also extract information about the signal damping from the time-dependent simulations. Tracing the height of the peaks as a function of distance or time, one can deduce the propagation length or the lifetime of the signals. Doing this for the fast signal, the same power law decay as observed in the frequency domain calculations was obtained, confirming that this signal indeed originates from the branch cut.
For the slower signal, we deduced the lifetime from the simulations as a function of the carrier frequency. The result is compared with the lifetime that was obtained from the imaginary part of the dispersion relation in Fig.~\ref{fig:pp_dispersionrelations}. The comparison is shown in Fig.~\ref{fig:pp_lifetimes}. It is obvious that the decay constant exactly matches $\textnormal{Im}[\omega]$, another confirmation that the slower signal is indeed due to the collective plasmons.

It is important to stress that, even though realistic dissipation was taken into account, the group velocity provides a good fit to the obtained signal velocities. Even for simulations with twice the realistic amount of ohmic losses (not shown), the match between both velocities still is very good.
This indicates that it is not the amount of dissipation that is important, but its variation with respect to the wavevector or frequency, i.e. the changes in the imaginary part of the dispersion relation. Close to the light line, the variation in both the real and imaginary part of the dispersion relation is large, indicating that in this region the group velocity is an ill-defined quantity.

\section{Branch cut contribution: precursor}
\label{ssc:pp_resultstimediscussion}
\begin{figure}[ht!]
\begin{center}
\includegraphics[width=\columnwidth]{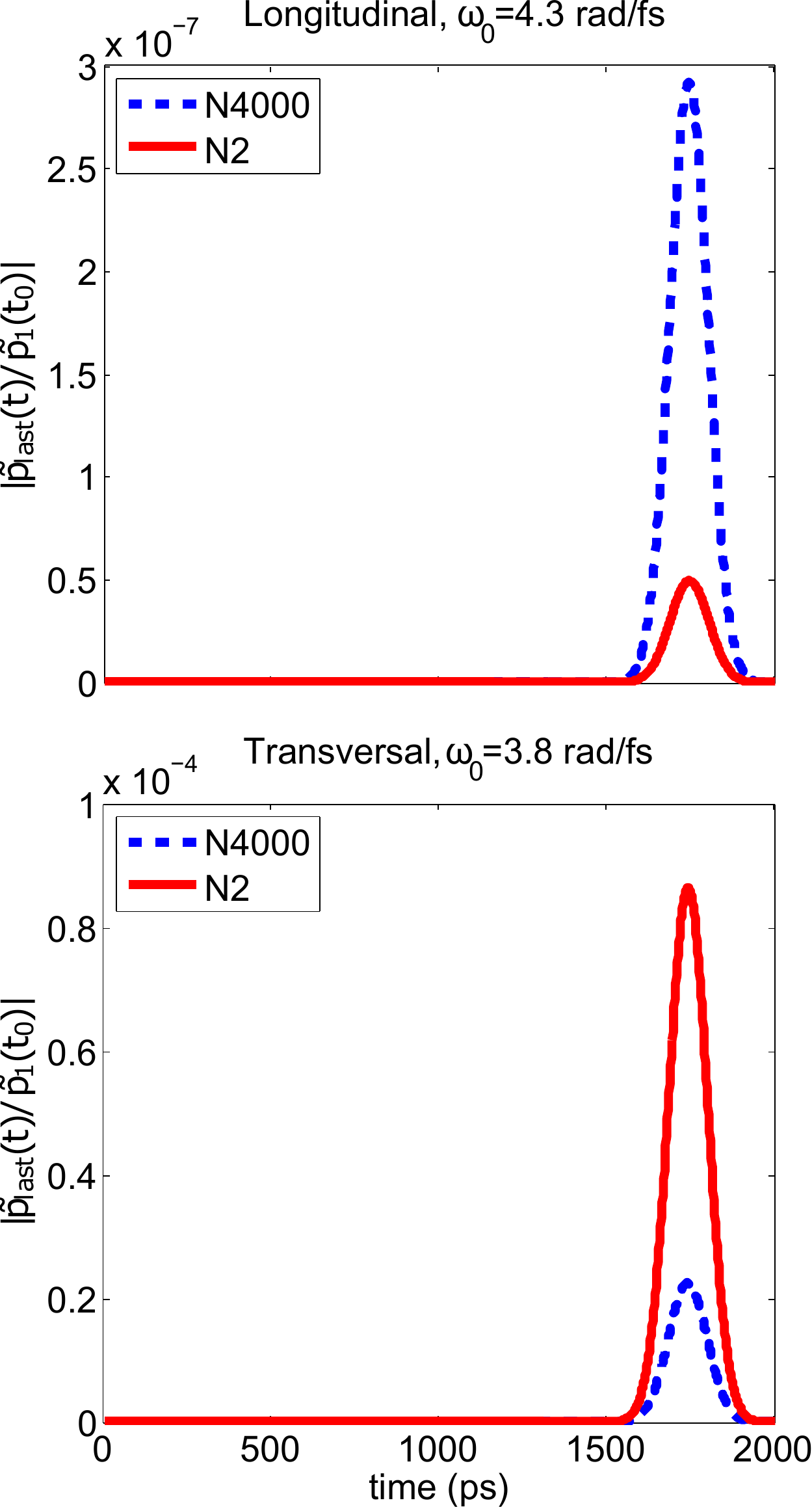}
\caption{\label{fig:two-particle_chain}
Same as in Fig.\ref{fig:pp_prop_contour}, only for an arrray comprizing two silver nanosphere, $n = 1$ and $n = 4000$, while removing all others in between by the ambient material.}
\end{center}
\end{figure}
As is seen from the analysis performed in the preceding section, the fast signal originates from the branch cut of the integrand in Eq.~(\ref{eq:pp_greensfunction}). Both the high propagation velocity and the slow geometrical decay make this signal very interesting from the viewpoint of fundamental physics as well as practical applications. The question that should be answered is how to interpret the branch cut contribution physically. First, we note the branch cut arises from the dipole sum, so that the fast signal is not simply the electric field that the first ($n = 1$) dipole generates at each particle in the chain, but rather reflects a collective effect, coming from the sum of all forward scattered fields.

To illustrate this, we carried out calculations for an array consisting of two silver spheres ($n = 1$ and $n = 4000$) in the dielectric background, separated by the same distance as in the original problem presented in the paper. We calculated the magnitude of the dipole moment on the last particle relative to the one on the first (excited) particle as a function of time around the time, when the precursor arrives there. The results are shown in Fig.~\ref{fig:two-particle_chain}. As is seen from the figure, this magnitude clearly differs from the one in the case where the full chain of particles is considered. In the case of longitudinal polarization, the magnitude of the last dipole is  enhanced by the intermediate particles, while for transverse polarization, it is reduced. The enhancement in the longitudinal case, we attribute to the fact that the near-field interactions between neighboring particles are relatively important then (since the radiative long-range interaction is absent here), while the reduction in the transverse case may be attributed to (distructive) interference effects between the secondary radiation fields (absent for longitudinal polarization) emitted by the $\lambda/2$ anti-phased domains of the chain. This proves that the precursor effect observed is not just free-space propagation of an electromagnetic field between the first and the last particle.

Since the electric field generated by an oscillating dipole propagates with the speed of light, it is understandable that such a signal should always be present. Its intensity is expected to maximize in the vicinity of the light line, where the frequencies of light $ck_b $ and collective plasmons $\omega(q)$ are equal to each other and also the wave vectors of light and plasmon modes match ($k_b = q$), ensuring the simultaneous occurrence of resonance and phase-matching between both types of modes. Away from the light line, the phase-matching condition is also fulfilled, however, the resonance condition is not (and {\it vise versa}). Therefore, the fast signal will also exist (see Fig.~\ref{fig:pp_prop_contour}), but less intense.

This is exactly what we observe from Fig.~\ref{fig:pp_freqdomain}. Indeed, for the longitudinal polarization, the transmitted signal acquires its maximum at $\omega = 4.12$ rad/fs, which corresponds to the crossing point of the corresponding dispersion curve with the light line (see Fig.~\ref{fig:pp_dispersionrelations}a). In the transversal case, this does not happen because the losses diverge at the crossing point (see Fig.~\ref{fig:pp_dispersionrelations}b), thus killing the signal at this frequency. Instead, the transversal signal is maximized at $\omega = 3.46$ rad/s, the point where the interplay between the resonance and phase-matching is optimal.

The fast signal resembles the so-called Sommerfeld-Brillouin forerunner.~\cite{Sommerfeld1914,Brillouin1914,Brillouin1960} Sommerfeld and Brillouin studied theoretically the propagation of a step-modulated light pulse in a Lorentz absorbing medium with a single broad absorption line. They showed that at propagation distances at which the medium is opaque in a broad spectral range (including the carrier frequency), the transmitted signal consists of two consecutive transients with essentially $100\%$ transmission, preceding the development of the steady-state response at the frequency of the incident field, expected from Beer-Lambert law. The transients have been named "forerunners". The faster signal (the "Sommerfeld forerunner") and the slower one (the "Brillouin forerunner") are formed, respectively, by frequencies large and small compared to the frequency of the absorption line: the spectral regions where the medium is somewhat transparent. The front of the former propagates at the velocity of light in vacuum,~\cite{Sommerfeld1914} while the latter moves approximately at the zero frequency group velocity.~\cite{Brillouin1914,Brillouin1960} After the seminal works of Sommerfeld and Brilluoin, the forerunners became a canonical problem of wave propagation in weakly dispersive absorbing media. A comprehensive bibliography on forerunners can be found in Ref.~\onlinecite{Oughstun2009}.

During the last decade, precursors have been observed in weakly dispersive dielectric media in regions of anomalous dispersion, using very narrow transitions of gases of cold potassium ($^{39}$K)~\cite{Jeong2006} and ($^{85}$Rb)~\cite{Wei2009} atoms, generated in a vapor-cell magneto-optic trap. Importantly, the forerunners have been found to possess an appreciable intensity, comparable with the main (carrier frequency) pulse.
At short distances, the Sommerfeld and Brillouin forerunners overlap and can not be distinguished; the common signal has been called "precursor",\cite{Jeong2006,Wei2009,Macke2013} the term that stands for this response in contemporary literature.

Turning to our case of LSP propagation, we emphasize that the conditions for the appearance of a precursor here differ substantially from the standard ones. First, the energy spectrum of the system under consideration is not a single line, but represents two highly dispersive branches. Second, we apply a Gaussian pulse, instead of a step-modulated one, having a spectrum that is narrow as compared to the systems's energy spectrum. In addition, for the set of parameters we used, the reduced wavelength $\lambda/2\pi$ of the incident field is on the order of the interparticle spacing $d$ ($k_p d \sim 1$), meaning that the system appears to be really a discrete medium; in this limit, the radiative interparticle interaction is important. Furthermore, the external field acts only on the leftmost particle, not propagating itself through the chain. Thus, the precursor under this type of excitation has its origin solely in the unidirectional propagation of LSPs, which travel with the speed of light and decrease geometrically, according to the decay law of the retarded interparticle dipole-dipole interaction. Nevertheless, in spite of these differences with the classical Sommerfeld-Brillouin forerunners, the fast signal found by us also bears the characteristics of a precursor.

\begin{figure}[t!]
\begin{center}
\includegraphics{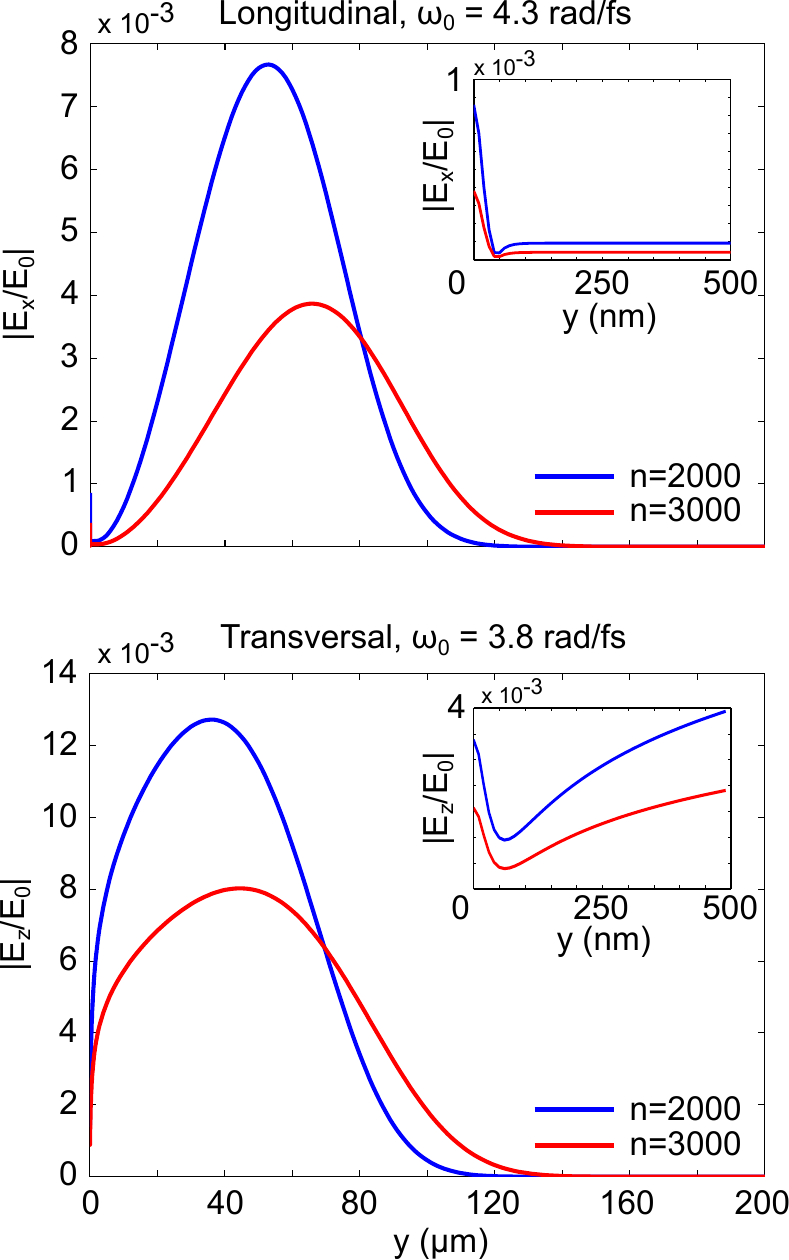}
\caption{\label{fig:pp_confinement} The electric field profile of the forward scattered signal is calculated along the line parallel to the y-axis situated at two positions in the chain: between the particles $n=2000$ and $n=2001$, i.e., the line $(x_{2000}+d/2,y,0)$, and similarly between particles $n=3000$ and $n=3001$. The excitation is the same as in Fig.~\ref{fig:pp_prop_contour} and the electric field is normalized with respect to the amplitude of the external excitation at $\omega_{0}$, $E_{0}$. The main figures give the response far away from the chain and the insets show the response close to the chain axis. Note that in the inset the distance is given in nm, rather than in $\mu$m as in the main plot.}
\end{center}
\end{figure}

\subsection{Lateral field profile of the precursor}
\label{confinement of the precursor}

Signal transmission through plasmonic arrays is studied mostly in the context of sub-diffraction waveguides. It is known that for such systems there exists a trade-off between confinement and losses:~\cite{Khurgin2015} having a truly sub-wavelength confinement, we lose in propagation length. In the present case, the collective plasmon signal dominates the transport at the beginning stage of the propagation; this contribution attenuates exponentially. However, after a certain distance, the decay shows a power-law behavior, indicating that the precursor comes into play. Due to the slow decay, it can propagate for remarkable distances.

Naturally, the question arises if this signal is also, in some sense, confined to the array. To answer this question, we calculated the electric field within a plane oriented perpendicular to the chain axis. Figure~\ref{fig:pp_confinement} shows the electric field components corresponding to the precursor signal as shown in Fig.~\ref{fig:pp_prop_contour}. The field is presented along two lines parallel to the y-axis, one located in between particles $n=2000$ and $n=2001$, and the other between $n=3000$ and $n=3001$. The main figures show the fields far away from the chain, the insets display the fields near to the chain (note the difference in units on the horizontal axis, in $\mu$m and nm, respectively). The figure clearly shows that, far away from the chain, the field intensity decreases and broadens, exactly as expected for a signal that decays geometrically. Interestingly, this behavior is not observed close to the chain. Even though the amplitude of the field decreases, it does not spread laterally. From this we can conclude that this part of the precursor is confined to the chain.

\section{Summary and Outlook}
\label{sc:pp_summary}
We have investigated theoretically the propagation of a localized surface plasmon through an array of equidistantly spaced spherical metal nanoparticles and have found that two different signals contribute to the optical response of the array: one traveling with the group velocity and attenuating exponentially and another one propagating with the velocity of light in the surrounding medium and decaying geometrically. We have shown that the former signal arises from the excitation of the array's normal modes, i.e. the collective plasmons of the chain of MNPs. The latter signal can not be deduced from the normal modes of the system. This contribution has been interpreted as a precursor - a unidirectional propagation of LSPs, traveling with the speed of light. As explained in Sec. V, the precursor can be interpreted as the multiple forward scattering of the field initially irradiated by the first particle ($n = 1$), which is excited by the external source. This indeed leads to a signal propagating with the background speed of light and the various phase relations between the fields emitted by the individual particles, resulting finally in the lateral confinement observed.

Although the precursor found in our study resembles the well known Sommerfeld-Brillouin forerunner, the conditions for appearance of this signal differ substantially from the standard situation: (i) - the medium we are dealing with is highly dispersive, (ii) - the discreteness of the system is of vital importance, (iii) - the external field has a narrow spectrum and acts only on the leftmost particle.

By calculating the electric field profile along a line perpendicular to the chain axis, the lateral dimensions of the forward scattered (precursor) signal have been studied. It has been shown that, surprisingly, even though the signal does not have sub-wavelength dimensions, close to the chain axis the profile does not change as a function of the propagation distance, indicating that this part of the signal is confined to the array.

The precursor dominates the LSP transmission at large distances, rendering this signal of fundamental and possibly practical importance to the signal propagation in plasmonic arrays. Although this signal is not fully confined to the array, a stronger confinement can be achieved by placing the array in a layered environment. This does not only lead to waveguiding of the signal, but may also affect the signal by changing the effective interactions between the various particles.~\cite{Compaijen2013}

To conclude, we notice that arrays comprizing dielectric nanoparticles with a high index of refraction (such as silicon) also support the subwavelength guiding for distances exceeding several tens of micrometers.~\cite{Savelev2014} Thus, it would be interesting for future studies to consider similar propagation effects in such arrays.

\acknowledgments

The authors acknowledge B.J. Hoenders and V.Y. Chernyak for fruitful discussions, and A. Malyshev, X. Inchausti, and M. Blok for contributing to the initial stage of this project.
This work was supported by NanoNextNL, a micro and nanotechnology consortium of the Government of the Netherlands and 130 partners.


\begin{thebibliography}{100}
\bibitem{footnote}
Srticly speaking, the Mie coefficient $a_1$ should be used instead of the polarizability given in Eq.~(\ref{eq:pp_polarizability}) The refinement would affect to some extent the propagation length of the chain's normal modes, but not the existence and geometrical decay of the precursor, because its nature has no relation to the polarizability.
\end{thebibliography}
\end{document}